\mag=\magstephalf 
\hsize=14.2 truecm \vsize=21 truecm
\newdimen\vu \vu=12 truebp 
\hoffset=6.5truein \advance\hoffset by -\hsize \hoffset=.5\hoffset
\parindent=2.5\vu 
\parskip 0pt plus .1\vu
\abovedisplayskip=.25\vu plus .2\vu minus .15\vu
\abovedisplayshortskip=0pt plus .2\vu
\belowdisplayskip=.25\vu plus .2\vu minus .15\vu
\belowdisplayshortskip=.15\vu plus .2\vu minus .1\vu
\normalbaselineskip=1.25\vu
\font\titl=cmbx10 scaled 1100 \font\tensc=cmcsc10
\font\ninerm=cmr9 \font\ninebf=cmbx9 \font\ninei=cmmi9
\font\nineit=cmti9 \font\ninesy=cmsy9
\font\ninesc=cmcsc10 at 9pt \font\ninett=cmtt9
\font\sixrm=cmr6 \font\sixi=cmmi6 \font\sixsy=cmsy6
\newfam\scfam \textfont\scfam=\tensc \def\sc{\fam\scfam\tensc}%
\def\ninept{\textfont0=\ninerm \textfont1=\ninei \textfont2=\ninesy
\scriptfont0=\sixrm \scriptfont1=\sixi \scriptfont2=\sixsy
\textfont\itfam=\nineit \textfont\bffam=\ninebf \textfont\ttfam=\ninett
\textfont\scfam=\ninesc
\def\rm{\fam0 \ninerm}%
\def\it{\fam\itfam\nineit}%
\def\bf{\fam\bffam\ninebf}%
\def\tt{\fam\ttfam\ninett}%
\def\sc{\fam\scfam\ninesc}%
\rm}

\newif\ifpagenos \pagenosfalse \pageno=-1
\output={\offinterlineskip\setbox0=
\vbox to 1\vu{\vfil\centerline{\head}}\dp0=0pt
\setbox1=\vbox to 1\vu{\vfil\centerline{\foot}}\dp1=0pt
\shipout\vbox{\box0\vskip1\vu \box255 \vskip1\vu\box1}
\global\advance\pageno by 1 \ifnum\pageno=0\global\pageno=1\fi}
\def\head{\ifnum\pageno<2 {}\else
\ifodd\pageno {\tensc Numerical Techniques $\ldots$}\else
{\tensc C. F. F. Karney}\fi\fi}
\def\foot{\ifnum\pageno<1 {}\else
\ifpagenos {\tenrm \the\pageno}\fi\fi}

\newif\ifadj
\def\ninedisp{\ninept\global\adjtrue \baselineskip=.9\vu
\everycr{\noalign{\ifadj \global\adjfalse
\vskip-\baselineskip\vskip\normalbaselineskip\fi}}}
\def\calcode{\catcode`|=4 \catcode`&=12
\setbox0=\hbox{$\it loop$}\setbox1=\hbox{$\it jam$}\setbox2=
\hbox{$\it s3\mathbin{{*}\rm r}v7$}\calcodea}
\def\calcodea#1{\ninedisp
\tabskip=2.5\vu\halign to \displaywidth
{\hbox to \wd0{$\it ##$\hfil}\tabskip 1.75\vu&\hbox to \wd1{$\it ##$\hfil}%
&\hbox to \wd2{$\it ##$\hfil}&\rm ##\hfil\tabskip\centering\crcr
#1\crcr}}
\def\timcode{\catcode`|=4 \catcode`&=12
\setbox0=\hbox{$\it loop$}\setbox1=\hbox{$\it jam$}\setbox2=
\hbox{$\it s3\mathbin{{*}\rm r}v7$}\timcodea}
\def\timcodea#1{\ninedisp
\tabskip=2.5\vu\halign to \displaywidth
{\hbox to \wd0{$\it ##$\hfil}\tabskip 1.75\vu&\hbox to \wd1{$\it ##$\hfil}%
&\hbox to \wd2{$\it ##$\hfil}&\hfil ##\tabskip1\vu
&\hfil\it ##&\hfil ##&\hfil ##&\hfil ##&\hfil ##&\hfil ##\tabskip\centering
\crcr
#1\crcr}}

\def\ifempty#1{\def\emptya{#1}\ifx\emptya\empty}
\newcount\refcnt \refcnt=1 
\def\qdef#1{\xdef#1{\number\refcnt}
	    \global\advance\refcnt by 1 }
\def\defrefs#1{\qqdef#1\enddefs}
\def\qqdef#1#2\enddefs{\qdef#1 \ifempty{#2}\else\qqdef#2\enddefs \fi}
\def\refno#1.  {\par\noindent\setbox0=\hbox{99.\ }\hangindent=\wd0
\hbox to \wd0{\hfil #1.\ }}

\defrefs{\channon \henon \chirikov \karney
\miller \rannou \knuth \metafont \minsky \cal \cft \nelson}

\newwrite\reffile \openout\reffile=berk.ref
\def\cite#1{\unskip\nobreak\relax\write\reffile{#1}%
    \def\tempa{$\mathsurround=0pt ^{\hbox{\the\scriptfont0 #1}}$}%
    \futurelet\tempc\citexx}
\def\citexx{\ifx.\tempc\let\tempd=\citepunct\else
    \ifx,\tempc\let\tempd=\citepunct\else
    \let\tempd=\tempa\fi\fi\tempd}
\def\citepunct{\tempc\edef\sf{\spacefactor=\the\spacefactor\relax}\tempa
    \sf\gobble}
\def\gobble#1{}

\def\section#1\par{\vskip 1.5\vu plus .2\vu
\vskip 0in plus .8in\penalty -50\vskip 0in plus -.8in
\noindent{\sc #1}\par\penalty 1000\vskip .5\vu plus .05\vu
\penalty 1000\noindent}

\def\references\par{\section References\par
\vskip\baselineskip\penalty 1000 \baselineskip=.9\vu
\vskip-\baselineskip\penalty 1000 \frenchspacing\ninept}

\def\eqn#1{$$\hbox to \displaywidth{\hskip\parindent
		$\displaystyle#1$\hfil}$$}
\def\eqno{\hfill\rm}
\topskip 4\vu plus .1\vu
\rightline{PPPL--2218 (1985)}
\vskip 3\vu
\centerline{\titl NUMERICAL TECHNIQUES FOR THE STUDY OF}
\vskip .3\vu
\centerline{\titl LONG-TIME CORRELATIONS}
\vskip 3\vu plus .1\vu
\centerline{\sc Charles F. F. Karney}
\centerline{Plasma Physics Laboratory, Princeton University, P.O. Box 451}
\centerline{Princeton, New Jersey 08544, U.S.A.}
\vfill
\noindent Presented at the Workshop on Orbital Dynamics and
Applications to Accelerators, Lawrence Berkeley Laboratory, Berkeley,
California, March 7--12, 1985.  This was published in Particle
Accelerators {\bf 19}(1--4), 213--221 (May 1986).
\eject
\baselineskip 1\vu
\par{\titl NUMERICAL TECHNIQUES FOR THE STUDY OF}
\vskip 1pt
\par{\titl LONG-TIME CORRELATIONS*}
\topskip 1\vu \vskip 3\vu plus .1\vu
\par{\sc Charles F. F. Karney}
\par Plasma Physics Laboratory, Princeton University, P.O. Box 451,
\par Princeton, New Jersey 08544, U.S.A.
\vskip 2\vu plus .1\vu\baselineskip=.9\vu
\hangindent=\parindent
	{\ninept{\it Abstract}\ \ \ \ In the study of long-time
correlations extremely long orbits must be calculated.  This may be
accomplished much more reliably using fixed-point arithmetic.  Use of
this arithmetic on the Cray--1 computer is illustrated.}

\baselineskip=\normalbaselineskip
\vskip1\vu plus .1\vu

\section Introduction

There has been considerable interest recently in simple dynamical
systems which exhibit very complicated behavior.  One of the
simplest such systems is a two-dimensional area-preserving map.  This
shares many of the properties of Hamiltonian systems with two degrees
of freedom.  The phase space for these systems is typically divided in
integrable and nonintegrable (or stochastic) portions.  An interesting
problem is the behavior of an orbit in the nonintegrable portion of
phase space when it approaches an integrable portion.  This can
introduce very long-time correlations into the stochastic orbits.  The
first systematic study of this problem was given by Channon and
Lebowitz \cite\channon, who studied orbits in the H\'enon map
\cite\henon.  This study was based on $7750$ orbits of length $10^4$.
However, subsequent studies have recognized that longer orbits must be
considered in order to determine the long-time behavior accurately.  For
example, in their work on the whisker mapping Chirikov and Shepelyansky
\cite{\chirikov} studied a single orbit of length $10^8$.  More ambitious
calculations were performed by the author \cite{\karney} on the
periodic quadratic mapping
	\eqn{Q:\qquad y'-y=g(x),\quad x'-x=y',\eqno(1)}
	where
	\eqn{g(x)=\cases{2(x^2-K),&
for $x_{\rm min}\le x<x_{\rm max}$,\cr
g(x\pm L),&otherwise,\cr}}
	and $L=x_{\rm max}-x_{\rm min}>0$.  In this study $1600$
orbits of length $2\times 10^9$ were used (i.e., a total of
$3.2\times10^{12}$ iterations).
\par\penalty1000\vskip .5\vu plus 1in\penalty1000
\hrule width 2truein\penalty1000
\vskip .5\vu plus .5in\penalty1000
{\ninept\baselineskip .9\vu\noindent
*This work was supported by the United States Department of Energy under
Contract DE--AC02--76--CHO--3073.
\par}\eject
\section The Numerical Problem

In a calculation of this magnitude, we must ask whether the results
obtained on a computer have any relevance to the study of the exact
system.  The problem arises because only a finite set of numbers can be
represented on a computer.  Consequently, {\it every} orbit in a
numerical mapping is periodic.  The numerical mapping does not give the
generic behavior for an exact mapping in which periodic orbits are a
set of measure zero.  Nevertheless, useful information can be obtained
if the average period $T$ of orbits in the numerical mapping exceeds
the time in which we are interested.  On the other hand, the numerical
calculations are useless if $T$ is less than the time in which we are
interested.

For a two-dimensional map, such as Eq.~(1) implemented in
floating-point arithmetic, we estimate that $T\sim 2^{p/2}$ where $p$
is the number of bits of precision.  On the Cray--1, we have $p=48$ and
$T\sim 10^7$.  Therefore, single-precision floating-point arithmetic on
the Cray--1 cannot be used to study orbits of length $10^9$.  One
possible solution (the brute-force approach) is to go to double
precision.  This extends $T$ to about $10^{14}$ but at a cost of a
factor of 2--4 in speed.  It is preferable, however, to understand what
defects in the floating-point number system cause $T$ to be
as short as it is, and then to employ a system of arithmetic which
doesn't have such defects.

We are approximating a mapping $Q$ (the ``exact'' mapping) with
another mapping $Q^*$ (the realization of $Q$ on a computer, the
``numerical'' mapping).  We can estimate the error in a single iteration
of the mapping as $2^{-p}$.  As we iterate the mapping, the error grows.
However, we can seek to control the error by making sure that $Q^*$ has
some of the same properties as $Q$.  Indeed if $Q^*$ has enough of the
``interesting'' properties of $Q$, we might tolerate quite a large
error in a single iteration.

(A parallel situation exists in the approximation of the collision
operator in a plasma by the Landau collision operator.  This is known to
be in error by about $5\%$.  However, because the Landau operator
conserves all the quantities conserved by the exact collision
operator---number, momentum, and energy---and because it has an $H$
theorem, we are sure that the errors won't affect anything
``important.''  Indeed, for these reasons, most plasma physicists are
happy to regard the Landau collision operator as exact!)

Perhaps the most important property of $Q$ is that it is
area-preserving.  It is known that a small amount of dissipation greatly
alters a mapping.  Now $Q^*$ is a mapping defined on a discrete set of
numbers so that the area-preserving property has to be translated into
the analogous property for discrete mappings.  However, because the
floating-point number system is nonuniform, this property is very
complicated and consequently difficult to implement.  On the other hand,
if a uniform number system is used, then area-preservation in $Q$
corresponds to the mapping $Q^*$ being {\it one-to-one}.  Such a number
system is implemented on most computers and is called the
``fixed-point'' number system.

\section Fixed-Point Numbers

Floating-point numbers are conventionally represented as $2^m\times f$
where ${1\over 2}\le f<1$ and $m$ can vary (the binary point can {\it
float}).  The fraction $f$ and the exponent $m$ are then stored in
different parts of the computer word.  Floating-point numbers are ideal
for representing numbers which may be very small or very large.  In typical
mapping calculations, this flexibility is not needed.  The numbers
representing the coordinates are usually bounded, so we do not need to
be able to represent very large numbers.  Furthermore, the added
precision available for small floating-point numbers is wasted since
these are often added to much larger numbers.

Fixed-point numbers are also represented as $2^m\times f$, but now
$0\le f<1$ and $m$ is {\it fixed}.  (Note a possible confusion in the
terminology:  {\it fixed-point} numbers have nothing to do with the {\it
fixed-points} of a mapping.)  Now only $f$ need be stored in the
computer word, and it is the programmer's responsibility to remember
$m$.  (Because a whole word is used to store $f$, it is often possible
to increase the precision.  Thus on a {\sc pdp}--10, 36 bits are
available for $f$, which is considerably more than the 27 bits used to
represent the fraction in floating-point numbers on that machine.)

Addition and subtraction of fixed-point numbers are exact.  Thus, if
Eq.~(1) is implemented in fixed-point arithmetic to give a numerical
mapping $Q^*$, then this can be represented by Eq.~(1) with the exact
$g(x)$ replaced by an approximation, $g^*(x)$.  Furthermore, $Q^{*-1}$
exists because the operations in $Q^*$ can be reversed to give $x$ and
$y$ in terms of $x'$ and $y'$.  In fact, there is a way to compute
$Q^{*-1}$ numerically without having to program the reverse
operations. If we define an involution ($J^2=$ identity)
	\eqn{J:\qquad y'=-y,\quad x'=x-y,}
	then we can show that $Q^{*-n}=J^{-1}Q^{*n}J$, which may be
exactly computed because $J$ can be exactly carried out in fixed-point
arithmetic.  Thus each point on the plane has a unique successor (given
by $Q^*$) and a unique predecessor (given by $Q^{*-1}$) and the mapping
is one-to-one.  Such a fixed-point mapping is a {\it permutation} on
phase space.  In contrast, a floating-point mapping is a {\it
many-to-one} mapping, or a {\it function} on phase space.

One-to-one mappings have been used in the study of dynamical systems by
Miller and Prendergast \cite{\miller} and by Rannou \cite{\rannou}.  These
authors implemented the mappings with integer arithmetic and used
rather coarse representations of phase space.  Thus Rannou divides phase
space into a maximum of $800\times 800$ cells.  (A fixed-point mapping
implemented on the Cray--1 has $2^{48}\times 2^{48}$ cells.)  However,
these studies are useful for providing theoretical results about
permutations.  In particular, Rannou \cite{\rannou} considers random
permutations of $N$ points defined as the ensemble of all possible such
permutations.  She shows that the average period of orbits is
${1\over2}(N+1)$.  The average period of orbits in a random function is
given by Knuth \cite{\knuth} as $\sqrt{\pi N/8} + {1\over 3}$.  The
reason that permutations have longer orbits is that they practice
collision-avoidance.  The only way such an orbit can become periodic is
by landing on the initial point.  In a random function, an orbit can
become periodic by landing on any of its previous points.

Now the mappings describing dynamical systems are definitely not
random.  In particular, they often possess symmetries (for example, the
symmetry defined by the involution $J$ connecting forwards and
backwards trajectories).  Rannou finds \cite{\rannou} that the average
length of the orbits of a random symmetric permutation is reduced to
$O(N^{1/2})$.  We conjecture that a similar phenomenon occurs with
symmetric functions reducing the average length of the orbits to
$O(N^{1/4})$.  Substituting $N=(2^{48})^2$, which is appropriate for a
two-dimensional mapping on the Cray--1, we find that $T\sim 10^{14}$
with fixed-point arithmetic and $T\sim 10^7$ with floating-point
arithmetic.  Clearly, fixed-point arithmetic allows us to examine much
longer orbits.

\section Operations on Floating-Point Numbers

Let us briefly describe how to perform some useful operations on
fixed-point numbers.  We begin with the elementary operations:  Addition
and subtraction are performed with the same instructions as for integer
addition and subtraction.  Multiplication by a power of two $2^{\pm m}x$
may be accomplished by a left/right shift of $x$ by $m$ bits (possibly
with sign-extension).  The computation of the integer part, $\mathop{\rm
int}(x)\equiv\lfloor x\rfloor$, and the fraction part, $\mathop{\rm
fract}(x)\equiv x-\lfloor x\rfloor$, of a number may be performed by
{\sc and}ing the computer word with appropriate masks.  (The computation
of $\mathop{\rm int}(x)$ for floating-point numbers needs a sequence of
instructions in {\sc fortran}:  $\it xi=\mathop{\rm aint}(x)$; $\it{\bf
if}\ (xi\mathbin{{.}\rm gt{.}}x)\ {\bf then}\penalty0\ xi=xi-\rm1.0$.)

The method for multiplying by an arbitrary number depends on the
computer.  For instance, on a {\sc pdp}--10 the {\sc mul} instruction
multiplies two 36-bit numbers to give the 72-bit product.  This result
can be shifted to align the binary-point with the assumed position
within the word.  Similarly, on the {\sc mc}68000 microprocessor the {\sc
muls} instruction gives the 32-bit product of two 16-bit quantities.
(Higher precision multiplication can be performed with a sequence of
these instructions.)  The situation is slightly different on the
Cray--1.  When two numbers with zero exponent fields are multiplied
using the floating-multiply instructions, the product of the fraction
fields is returned with no normalization performed.  Thus, if we
represent fixed-point numbers as a Cray--1 word with the binary point
48 places from the right, then we can multiply two numbers in $[0,1)$
with the rounded floating-multiply instruction ${*}\rm r$.

Special functions may be calculated by converting the fixed-point
number to floating-point, calling a library routine for the special
function, and converting the result back to fixed-point.  Alternatively,
a subroutine calculating the special function directly with fixed-point
arithmetic may be written.  The program {\sc metafont} by Knuth
\cite{\metafont} contains a complete collection of routines for the
elementary functions assuming a fraction size of 16 bits.  An
interesting shortcut is available for the calculation of random
numbers.  Usually, there is a library routine which returns a random
floating-point number uniformly distributed in $[0,1)$.  The fraction
field of this floating-point number is uniformly distributed in
$[{1\over 2},1)$; thus a uniformly distributed fixed-point number may
be obtained by taking all but the first bit of the fraction.

\section Examples of Mappings

Care must be taken when implementing an area-preserving mapping in
fixed-point arithmetic to ensure that the numerical mapping is
one-to-one. We have seen that the mapping $Q$ may be implemented in a
straightforward way.  What about other mappings?  The problem is well
illustrated by the mapping
	$x'=2x$, $y'=\textstyle{1\over 2}y$.
	If this is coded as it stands, then on each iteration the least
significant bit of $y$ is lost and the mapping is clearly not
invertible.  Some way is therefore needed for remembering that lost
bit.

We begin by observing that a succession of nonlinear shifts
	\eqn{x'=x+f(y),\quad y'=y+g(x')}
	is in a form that is invertible.  The trick is to combine
mappings of this form to produce the desired mapping.  Thus the 
scaling mapping
	\eqn{x'=sx,\quad y'=y/s}
	may be implemented as
	\eqn{x^*=x-y/s,\quad y^*=y+(s-1)x^*,
\quad x'=x^*+y^*,\quad y'=y^*-(s-1)x'/s.}
Similarly, the rotation
	\eqn{x'=\cos\theta\,x-\sin\theta\,y,\quad 
y'=\sin\theta\,x+\cos\theta\,y}
	can be coded as
	\eqn{x^*=x-\alpha y,\quad y'=y+\beta x^*,\quad x'=x^*-\alpha y',}
	where $\alpha =(1-\cos\theta)/\sin\theta$ and
$\beta=\sin\theta$.  Note that $\alpha$ diverges for
$\theta\rightarrow\pi$.  But this is no serious limitation because
rotations by multiples of ${1\over 2}\pi$ can be realized exactly
merely with sign changes;  thus we can restrict
$\left|\theta\right|\le{1\over 4}\pi$.  Interestingly, this form for
the rotation involves only three multiplications, and so may be faster
than the ``standard'' form which involves four.  If $\theta$ is small,
then the rotation can be approximated by \cite{\minsky}
	\eqn{x'=x-\epsilon y,\quad y'=y+\epsilon x',}
	where $\epsilon=2\sin{1\over 2}\theta$.  In fact, with exact
arithmetic this produces a slightly eccentric ellipse $x^2-\epsilon
xy+y^2=\rm const$.

\section Realization on a Cray--1

The biggest disincentive to using fixed-point arithmetic is the absence
of any support for this arithmetic in languages like {\sc fortran}.
Usually, one has to resort to assembly language to utilize this
arithmetic.  However, it is possible to make a substantial gain in speed by
coding in assembly language, so the effort is often worth it.  (Mapping
calculations tend to benefit from hand-coding in assembly language
because most of the running time is spent in a small mapping
subroutine.)  In order to illustrate the benefits, we show how the
mapping $Q$ may be written in assembly language {\sc cal} on the
Cray--1.  The reader is referred to the {\sc cal} manual \cite{\cal} for
further details on the language, and to the {\sc cft} manual
\cite{\cft} (Appendix F) for details on how to interface assembly
language routines to a {\sc fortran} program.

Fixed-point numbers are represented as a 64-bit word in twos-complement
notation with the binary point 48 places from the left.  The
floating-point multiply instruction ${*}\rm r$ can multiply two such
fixed-point numbers provided they lie in the range $[0,1)$.  It is
convenient to scale the variables in Eq.~(1) so that the mapping is
periodic in $x$ and $y$ with period $1$.  The running time is slightly
shortened if the inverse of the mapping is used.  The mapping then
becomes \cite\karney
	\eqn{Q^{*-1}:\qquad
x'=x-y+{\textstyle{1\over 2}}\pmod1,\quad
y'=y-2^m(ax^{\prime2}-bx'+c)\pmod1,}
	where $a$, $b$, $c$ are all in $[0,1)$ and $m$ is non-negative.
In addition, we wish to keep track of when $x$ or $y$ leave the unit
square.  The speed can be increased by following 64 particles at once (to
make use of the vectorization capabilities of the Cray--1) and by
iterating the mapping many times in one call.  Thus a {\sc fortran}
realization of this procedure (using floating-point arithmetic) reads
$$\ninedisp\setbox0=\hbox{99999}\tabskip1.5\vu
\halign to \displaywidth{\hbox to \wd0
{\hfil \it #}\tabskip 1\vu&$ #$\hfil\tabskip\centering\cr
&{\bf subroutine}\ \mathop{\rm quade}(n,x,y,lv,a,b,c,m)\cr
&{\bf parameter}\ (l=64)\cr
&{\bf logical}\ lv\cr
&{\bf dimension}\ x(l),y(l),xi(l),yi(l),lv(l)\cr
&{\bf do}\ {\it 1}\ i=1,l\cr
&\ \ lv(i)={.}{\rm false}{.}\cr
1&{\bf continue}\cr
&{\bf do}\ {\it 2}\ j=1,n\cr
&\ \ {\bf do}\ {\it 2}\ i=1,l\cr
&\ \ \ \ x(i)\phantom i=x(i)-y(i)+0.5\cr
&\ \ \ \ xi(i)=\mathop{\rm aint}\left(x(i)\right)\cr
&\ \ \ \ xi(i)=\mathop{\rm cvmgm}\left(xi(i)-1.0,xi(i),x(i)\right)\cr
&\ \ \ \ x(i)\phantom i=x(i)-xi(i)\cr
&\ \ \ \ y(i)\phantom i=y(i)-2.0\mathbin{{*}{*}}m*\left(a*x(i)\mathbin{{*}{*}}2-b*x(i)+c\right)\cr
&\ \ \ \ yi(i)=\mathop{\rm aint}\left(y(i)\right)\cr
&\ \ \ \ yi(i)=\mathop{\rm cvmgm}\left(yi(i)-1.0,yi(i),y(i)\right)\cr
&\ \ \ \ y(i)\phantom i=y(i)-yi(i)\cr
&\ \ \ \ lv(i)=lv(i)\mathbin{{.}{\rm or}{.}}
\left(xi(i)\mathbin{{.}{\rm ne}{.}}0.0\right)\mathbin{{.}{\rm or}{.}}
\left(yi(i)\mathbin{{.}{\rm ne}{.}}0.0\right)\cr
2&{\bf continue}\cr
&{\bf return}\cr
&{\bf end}\cr}$$
	This iterates $Q^{*-1}$ $n$ times, and returns $x$ and $y$.
(The {\sc cft} function $\mathop{\rm cvmgm}(a,b,c)$ returns $a$ if
$c<0$ and $b$ otherwise.)  The variable {\it lv} returns {\sc true} if
the particle left the unit square during any of these $n$ iterations.

Let us see how this can be coded using fixed-point arithmetic in {\sc
cal}.  For brevity, the loading and storing of the arguments is omitted.
At this point we have stored the vector length (64) in the register
$\it vl$, the vectors $x$ and $y$ in the vector registers $\it v2$ and
$\it v1$, the constants $a$, $b$, and $c$ in $\it s2$, $\it s3$, and $\it
s4$,  the count $n$ in $\it a1$, and the shift $m$ in $\it a4$.  We
begin with some initialization:
	$$\calcode{
|v0|\rm 0|${\it v0}\leftarrow0$ (used for {\sc or}ing with $x$ and $y$)\cr
|s1|\rm 1\cr
|s1|s1\mathbin<\rm 47|$0.5$\cr
|s6|-s1|${\it s6}\leftarrow-0.5$\cr
|s5|\mathbin<\rm 48|${\it s5}\leftarrow\hbox{fraction mask}$\cr
|a1|-a1|${\it a1}\leftarrow-n$ (flip sign of count)\cr}$$
	We next turn to the main loop.  Only the $j$ loop in the {\sc
fortran} code needs to be explicitly written.  The $i$ loop is
implicitly performed by the vector instructions.  The calculation is
carried out entirely in registers.  However, at the end of one
iteration, $x$ and $y$ are in different registers ($\it v7$ and $\it
v4$).  It is therefore necessary to repeat the coding with interchanged
registers to get $x$ and $y$ back to $\it v2$ and $\it v1$.  The test
for the particle having left the unit square $\it lv$ is given by {\sc
or}ing the successive $x$'s and $y$'s together into register $\it v0$.  At
the end we check whether the integer part of $\it v0$ is nonzero.  We
are able to postpone the operation $y\leftarrow\mathop{\rm fract}(y)$
until the end.  (With floating-point arithmetic, this would result in a
loss of precision, so it is necessary to extract the fractional part
with each iteration.)
	$$\calcode{
loop|v3|s6+v1|$y-0.5$\cr
|v5|v2-v3|$x\leftarrow x-y+0.5$\cr
|v7|s5\mathbin&v5|$x\leftarrow\mathop{\rm fract}(x)$\cr
|v6|s2\mathbin{{*}\rm r}v7|$ax$\cr
|v2|v0\mathbin!v5|${\it lv}\leftarrow{\it lv}\mathbin{\sc or}
\mathop{\rm int}(x)\ne0$\cr
|v3|v6\mathbin{{*}\rm r}v7|$ax^2$\cr
|v4|s4+v3|$ax^2+c$\cr
|s0|s3\mathbin{{*}\rm r}s4|use multiply unit for 1 cycle\cr
|v0|s3\mathbin{{*}\rm r}v7|$bx$\cr
|v6|v4-v0|$(ax^2-bx+c)$\cr
|v5|v6\mathbin<a4|$2^m(ax^2-bx+c)$\cr
|v4|v1-v5|$y\leftarrow y-2^m(ax^2-bx+c)$\cr
|v0|v2\mathbin!v4|${\it lv}\leftarrow{\it lv}\mathbin{\sc or}
\mathop{\rm int}(y)\ne0$\cr
|a1|a1+\rm 1|increment loop counter\cr
|v3|s6+v4|now repeat everything with\cr
|v5|v7-v3|\quad${\it v1} \rightleftharpoons {\it v4}$ and
${\it v2}\rightleftharpoons {\it v7}$\cr
|\ldots\cr
|a1|a1+\rm 1\cr
|a0|a1\cr
|{\rm jam}|loop|jump back if more to do\cr}$$
	We end by taking the fractional part of $y$ and by testing $\it
v0$ for a nonzero integer part.
	$$\calcode{
|v3|s5\mathbin&v1|$y\leftarrow\mathop{\rm fract}(y)$\cr
|a3|\rm 48|fraction shift\cr
|v6|v0\mathbin>a3|shift out fraction part of {\it v0}\cr
|vm|v6,n|${\it lv}\leftarrow
\mathop{\rm int}(y)\ne0 \mathbin{\sc or}\mathop{\rm int}(x)\ne0$\cr
|s1|vm|transfer to {\it s1}\cr}$$
	At this point $x$ and $y$ are in $\it v2$ and $\it v3$, and
$\it lv$ is in $\it s1$.

There are two sources of speed on the Cray--1.  The first is the ability
to process vectors.  This enables a given functional unit to produce one
result every cycle ($12.5\,\rm ns$).  Utilizing this feature in {\sc
cal} is relatively easy.  The second source of speed is the ability of
different functional units to be operating at the same time.  Depending
on the degree of overlap, this can speed up the program by a factor of
1.5--3.  However, taking advantage of this feature is made difficult by a
complicated set of rules for when a particular instruction can issue.
An extremely useful tool is the timing code {\sc cycles} \cite{\nelson}
which produces a detailed timing analysis of the code.  The application
of this code to the main loop of the mapping routine gives:
	$$\timcode{|||W|\rm D|I|C|O|F|R\cr
loop|v3|s6+v1|||0|5|64|68|69\cr
|v5|v2-v3|68|25|69|74|133|137|138\cr
|v7|s5\mathbin&v5|4|10|74|78|138|142|142\cr
|v6|s2\mathbin{{*}\rm r}v7|3|10|78|87|142|146|151\cr
|v2|v0\mathbin!v5|63|07|142|146|206|210|210\cr
|v3|v6\mathbin{{*}\rm r}v7|8|25|151|160|215|219|224\cr
|v4|s4+v3|8|10|160|165|224|228|229\cr
|s0|s3\mathbin{{*}\rm r}s4|58|01|219|226\cr
|v0|s3\mathbin{{*}\rm r}v7|||220|229|284|288|293\cr
|v6|v4-v0|8|10|229|234|293|297|298\cr
|v5|v6\mathbin<a4|4|10|234|240|298|302|304\cr
|v4|v1-v5|69|27|304|309|368|372|373\cr
|v0|v2\mathbin!v4|4|10|309|313|373|377|377\cr
|a1|a1+\rm 1|||310|312\cr
|v3|s6+v4|62|05|373|378|437|441|442\cr}$$
	\fontdimen8\tenex=.6pt For each instruction is given the
\underbar wait time, the \underbar delay code, the \underbar issue
time, the \underbar chain slot time, the \underbar operand ready time,
the \underbar functional unit ready time, and the \underbar result
ready time\fontdimen8\tenex=.4pt.  The times are all in units of the
clock of the Cray--1 namely $12.5\,\rm ns$.  The delay code is an octal
number describing why the instruction could not issue.  The meanings of
the bits in this code are
	$$\ninedisp\tabskip 2.5\vu
\halign to\displaywidth{\hfil \it
#\quad\tabskip0pt&#\hfil\tabskip\centering\cr
1&functional unit not ready\cr
2&result register not ready\cr
4&operand register not ready\cr
10&waiting for chain slot\cr
20&missed chain slot\cr}$$
	From the last line in the timing analysis, we see that one
complete iteration (for 64 particles) takes 373 cycles or $73\,\rm
ns/particle$.  Since there are 12 vector instructions in one iteration,
the machine is computing results at the rate of about 2/cycle.  Very
little improvement is possible beyond this because the vector add unit
is busy $91\%$ of the time.  For comparison, if the same mapping is
implemented with floating-point in {\sc fortran}, it takes $410\,\rm
ns/particle/iteration$, a factor of 5.6 slower.

The complexity of timing on the Cray--1 can be understood by
considering the ``dummy'' scalar multiply which issues at 219.  This
produces no useful result, but causes the next vector multiply
instruction to issue one cycle later.  Because of this, the vector add
unit is ready at the chain slot time for this instruction (229) and the
next instruction chains to this one.  Without this dummy instruction, the
vector multiply would issue at 219, the vector add would then miss the
chain slot and have to wait until time 292 to issue; i.e., there would
have been an additional delay of 63 cycles.

\section Conclusions

In this paper, we have shown that the area-preserving nature of
mappings implemented on a computer can be preserved by using
fixed-point arithmetic.  Because of this, much longer orbits can be
studied.  The precision is comparable to that of floating-point numbers;
however, round-off error is much easier to control with fixed-point
arithmetic.  At present, access to fixed-point numbers is through
assembly language.  However, it is often faster than floating-point
arithmetic.  Indeed, since floating-point instructions are not available
on several modern microprocessors, fixed-point arithmetic would be a
natural way of studying mappings on such devices.

\references

\refno \channon.  S. R. Channon and J. L. Lebowitz,
in {\it Nonlinear Dynamics}, Annals of the New York Academy of Sciences
{\bf 357}, 108 (New York, 1980); S. R. Channon, {\it Stochasticity and
Homoclinic Oscillations in an Area-Preserving Mapping}, Doctoral Thesis,
Rutgers University (1981).
\refno \henon.  M. H\'enon, {\it Quarterly of App. Math.} {\bf 27}, 291 (1969).
\refno \chirikov.  B. V. Chirikov and D. L. Shepelyansky, {\it Physica}
{\bf 13D}, 395 (1984).
\refno \karney.  C. F. F. Karney, {\it Physica} {\bf 8D}, 360 (1983).
\refno \miller.  R. H. Miller and K. H. Prendergast, {\it Astrophysical J.}
{\bf 151}, 699 (1968).
\refno \rannou.  F. Rannou, {\it Astron. and Astrophys.} {\bf 31}, 289 (1974)
and
{\it Etude Num\'erique de Transformations Planes Discr\`etes},
Doctoral Thesis, University of Nice (1972).
\refno \knuth.  D. E. Knuth, {\it The Art of Computer Programming}, 2nd
Edition, Vol. 2 (Addison Wesley, 1981).
\refno \metafont.  D. E. Knuth, {\it The METAFONTbook} (Addison Wesley,
1985).
\refno \minsky.  M. Minsky, in {\it HAKMEM}, Artifical Intelligence
Laboratory Report AIM--239, Mas\-sa\-chu\-setts Institute of
Technology (1972), item 149.
\refno \cal.  {\it Cray Assembly Language Reference Manual}
(Cray Research, Inc., 1980).
\refno \cft.  {\it CFT, the Cray--1 FORTRAN Compiler}, Publication
SR--0009 (Cray Research, Inc., 1984).
\refno \nelson.  H. L. Nelson, {\it Timing Codes on the Cray--1:
Principles and Applications}, Lawrence Livermore Laboratory Report
UCID--30179, Revision 2 (1981).

\bye